\documentstyle[prl,aps,multicol,epsf]{revtex}
\begin{document}
\draft
\widetext
\title{From the double-exchange Hamiltonian to the  $t-J$ model: Classical
spins}
\author{Eugene Kogan$^{1}$ and Mark Auslender$^2$}
\address{$^1$ Jack and Pearl Resnick Institute 
of Advanced Technology,
Department of Physics, Bar-Ilan University, Ramat-Gan 52900, 
Israel\\
$^2$ Department of Electrical and Computer Engineering,
Ben-Gurion University of the Negev,
P.O.B. 653, Beer-Sheva, 84105 Israel}
\date{\today}
\maketitle
\begin{abstract}
\leftskip 54.8pt
\rightskip 54.8pt
From the double-exchange Hamiltonian with
classical localized spins  in the limit of large but finit Hund
exchange coupling we obtain the $t-J$ model (with
classical localized spins).
\end{abstract}
\begin{multicols}{2}
\narrowtext

The Hamiltonian  of the DE model \cite{zener,anderson,degennes} is 
\begin{eqnarray}
\label{hamiltonian}   
H=-J_H \sum_{n\alpha\beta} {\bf m}_n\cdot 
{\bf \sigma}_{\alpha\beta}c_{n\alpha}^{\dagger} c_{n\beta}-\sum_{nn'\alpha}
t_{nn'} c_{n\alpha}^{\dagger} c_{n\alpha},
\end{eqnarray}
where $t_{n-n'}$ is the electron hopping,  $J_H$ is the effective Hund
exchange 
coupling between a core spin and a conduction electron,
$\hat{\bf \sigma}$ is the vector of the Pauli matrices, and $\alpha,\beta$ are
spin indices. We express the localized (classical) spin  by
a unit vector whose orientation is determined by polar angle $\theta$ and
azimuthal angle $\phi$. In a single electron representation the Hamiltonian 
 can be presented as
\begin{eqnarray}
\label{hamiltonian2}  
H_{nn'} = H^{ex}+H^{kin}= -J_H {\bf m}_n\cdot {\bf \sigma}\delta_{nn'}- t_{nn'}.
\end{eqnarray}
We consider the case of strong exchange: $J_H\gg W$ where $W$ is the electron band
width. In this case we should fist diagonalize  the exchange part of the
Hamiltonian. This is done by choosing
local spin quantization axis on each site in the direction of ${\bf m}$. 
In this
representation Hamiltonian  is \cite{kogan}
\begin{eqnarray}
\label{hamlocquant}  
H_{nn'} =  -J_H\sigma^z\delta_{nn'}
-t_{nn'}\left(\begin{array}{cc}a_{nn'} & b_{nn'}\\
b_{n'n}^* & a_{n'n}\end{array}\right),
\end{eqnarray}
where
\begin{eqnarray}
a_{nn'} =  \cos\frac{\theta_n}{2}\cos\frac{\theta_{n'}}{2}
+\sin\frac{\theta_n}{2}\sin\frac{\theta_{n'}}{2}e^{i(\phi_{n'}-\phi_{n})}\nonumber\\
b_{nn'}=\sin\frac{\theta_n}{2}\cos\frac{\theta_{n'}}{2}e^{-i\phi_{n}}-
\cos\frac{\theta_{n}}{2} \sin\frac{\theta_{n'}}{2}e^{-i\phi_{n'}}.
\end{eqnarray}
The transformation to local spin  quantization axis including an additional
Euler rotation angle, which leads to a more involved effective Hamiltonian than 
\ref{hamlocquant},  was introduced by Nagaev \cite{nagaev}.

The next step, like it is
done in the derivation of the $t-J$ model from the Hubbard model \cite{izyumov},
is to apply a canonical transformation 
\begin{eqnarray}
H\rightarrow \tilde{H}=e^SHe^{-S}=H+[S,H]+\frac{1}{2}[S[S,H]]+\dots
\end{eqnarray}
which excludes 
all band-to-band transitions. This can be achieved if we chose the operator $S$
in the form
\begin{eqnarray}
S_{nn'}=\frac{t_{nn'}}{2J_H}\left(\begin{array}{cc}a_{nn'} & b_{nn'}\\
-b_{n'n}^* & a_{n'n}\end{array}\right).
\end{eqnarray}
We have
\begin{eqnarray}
[S,H^{ex}]_{nn'}=t_{nn'}\left(\begin{array}{cc}0 & b_{nn'}\\
b_{n'n}^* & 0\end{array}\right)\nonumber\\
\left[S,[S,H^{ex}]\right]_{nn}
=2J_{nn'}\left(\begin{array}{cc}
 -|b_{nn'}|^2 & 0\\
0 & |b_{nn'}|^2\end{array}\right),
\end{eqnarray}
where $J_{nn'}=|t_{nn'}|^2/(2J_H)$.
Keeping terms up to the second order with respect to $t$ (and only site-diagonal
part of the second order terms) we obtain 
\begin{eqnarray}
\label{hamlo}  
\tilde{H}_{nn'} =  -J_H\sigma^z\delta_{nn'}
-t_{nn'}\left(\begin{array}{cc}a_{nn'} &0\\
0 & a_{n'n}\end{array}\right)\\
+\sum_{n''}J_{nn''}({\bf m}_n\cdot{\bf m}_{n''}-1)\sigma^z\delta_{nn'},
\end{eqnarray}

In the second quantization form the Hamiltonian (\ref{hamlo})  has the form
(ignoring the constant term)
\begin{eqnarray}
\label{ham}
\tilde{H}=-\sum_{nn'}t_{nn'}a_{n'n}d_n^{\dagger}d_{n'}+\sum_{nn'}J_{nn'}{\bf
m}_n\cdot{\bf m}_{n'}(1-d_n^{\dagger}d_{n}),
\end{eqnarray}
where for the case of
hole doping $d^{\dagger}(d)$ is the operator of creation (annihilation) 
of the hole, and for the case of electron doping
it is the operator of creation (annihilation) of the electron.

Looking at the  Hamiltonian obtained, we see that large but finite
Hund exchange dynamically generates antiferromagnetic exchange 
(the second term in Eq. (\ref{ham}). This term, however, 
is not independent upon the electron (hole) subsystem. In this regard this
Heisenberg like term resembles the first non-Heisenbergian term (kinetic
exchange).

\end{multicols}
\end{document}